# A more direct representation for complex relativity

**D.H. Delphenich**[*]

Physics Department, Bethany College, Lindsborg, KS 67456, USA

An alternative to the representation of complex relativity by self-dual complex 2-forms on the spacetime manifold is presented by assuming that the bundle of real 2-forms is given an almost-complex structure. From this, one can define a complex orthogonal structure on the bundle of 2-forms, which results in a more direct representation of the complex orthogonal group in three complex dimensions. The geometrical foundations of general relativity are then presented in terms of the bundle of oriented complex orthogonal 3-frames on the bundle of 2-forms in a manner that essentially parallels their construction in terms of self-dual complex 2-forms. It is shown that one can still discuss the Debever-Penrose classification of the Riemannian curvature tensor in terms of the representation presented here.

**Contents**

## 1 Introduction

At the root of the methods of complex relativity [1-5] is the fact that the Lorentz group $O(3, 1)$ – or rather, its Lie algebra $\mathfrak{so}(3, 1)$ – admits various isomorphic representations. Perhaps the most studied representation is the (1-to-2) representation of the proper orthochronous Lorentz group $SO_0(3, 1)$ by the elements of the group $SL(2; \mathbf{C})$, which leads one into the realm of Dirac spinors and relativistic wave mechanics. One approach to complex relativity then involves the representation of 2-forms on spacetime by $SL(2;\mathbf{C})$ spinors (see Obukhov and Tertychniy [5]). Another approach to complex relativity is based in the fact that there is also an isomorphism of $SO_0(3, 1)$ with the complex orthogonal group in three dimensions $SO(3; \mathbf{C})$, which acts naturally on $\mathbf{C}^3$.

---

[*] E-mail: david_delphenich@yahoo.com

The customary way of representing $SO(3; \mathbf{C})$ in complex relativity is by way of self-dual complex 3-frames that are associated with 2-forms on spacetime. In order to make this association, one complexifies the six-real-dimensional vector space of 2-forms on Minkowski space into a six-complex-dimensional vector space and then restricts to the three-complex-dimensional subspace of self-dual complex 2-forms, which satisfy the eigenvalue equation:

$$*F = iF \qquad (1.1)$$

relative to the Hodge * operator that one defines on the bundle $\Lambda^2(M)$ of 2-forms on the spacetime manifold $M$ in terms of the Lorentzian metric $g$ and a Lorentzian unit-volume element $V$, and extends to the complexification of $\Lambda^2(M)$.

The basic thesis of this presentation is that since * will define an almost-complex structure on the real vector bundle $\Lambda^2(M)$ to begin with, there is something redundant about complexifying the bundle, as well, when one is only going to use a rank three sub-bundle for representing Lorentz transformations. Apparently the original desire was to adapt the methods of self-duality that had proved so fundamental to the study of gauge fields on Riemannian manifolds, in which the restriction of * to 2-forms obeys $*^2 = I$, which makes the eigenvalues of * real, to the Lorentzian case, in which $*^2 = -I$, which makes them imaginary.

The key is to regard (1.1), not as an eigenvalue *equation*, but a defining *identity*. In particular, it facilitates the definition of the action of complex scalars on the fibers of $\Lambda^2(M)$ that makes it into a complex vector bundle of rank three. The isomorphism of a fiber of $\Lambda^2(M)$ with $\mathbf{C}^3$ then follows from making a choice of complex 3-frame for the fiber. Of course, unless $M$ is parallelizable a unique choice of complex 3-frame for each fiber is only locally possible.

In order to show that this produces a more direct route from 2-forms to complex 3-vectors that would be of interest to relativity, we shall demonstrate how one represents the various results of complex relativity. In the first section, we summarize the definition of the complex orthogonal group for three dimensions and exhibit its isomorphism with the identity component of the Lorentz group. Next, we discuss the representation of 2-planes in $\mathbf{R}^4$ by decomposable bivectors and algebraic 2-forms, the complexification of these spaces, and some of the real and complex scalar products of physical significance that facilitate the representation of $SO(3; \mathbf{C})$ in those spaces. We also show how the methods that are discussed in the present work relate to the more established methods of complex relativity. Then, in order to show how the representation of the geometric machinery of general relativity can be independently formulated on the bundle of oriented complex 3-frames in the bundle of 2-forms on spacetime, we first summarize the conventional representation of general relativity in terms of real frames in the tangent bundle in such a manner that the construction in the complex case is completely analogous. We conclude by discussing some topics of research that would make the analogy complete.

## 2 The complex orthogonal group

Although one is often taught in special relativity that the vector product that one learned about in the context of the three-dimensional real vector space $\mathbf{R}^3$ is no longer applicable to the four-dimensional real vector space $\mathbf{R}^4$, it is interesting that actually all one has to do to make the cross product relativistically significant is to complexify it. This is because on $\mathbf{R}^3$ the cross product defines a Lie algebra that is isomorphic to $\mathfrak{so}(3; \mathbf{R})$ by way of:

$$[\mathbf{v}, \mathbf{w}] = \mathbf{v} \times \mathbf{w} . \tag{2.1}$$

If $\mathbf{e}_i$, $i = 1, 2, 3$ are the standard basis vectors for $\mathbf{R}^3$ – viz., the triples real numbers of the form $[\delta_i^1, \delta_i^2, \delta_i^3]$ – then the structure constants of the Lie algebra relative to this basis are given by the Levi-Civita symbol $\varepsilon_{ijk}$, since, as one directly verifies:

$$[\mathbf{e}_i, \mathbf{e}_j] = \varepsilon_{ijk}\, \mathbf{e}_k . \tag{2.2}$$

This makes the components of $\mathbf{v} \times \mathbf{w}$ relative to this basis equal to:

$$(\mathbf{v} \times \mathbf{w})_i = \varepsilon_{ijk} (v_j w_k - v_k w_j) . \tag{2.3}$$

In the case of $\mathbf{C}^3$, if associates the basis vectors $\mathbf{e}_i$ with the same vectors, regarded as complex vectors with vanishing imaginary parts, then one can still define the cross product of two vectors by (2.3), and the result is that it defines a Lie algebra on $\mathbb{C}^3$ by way of (2.1), whose structure constants are still given by (2.2). However, the Lie algebra is now isomorphic to $\mathfrak{so}(3; \mathbf{C})$.

As it turns out, this Lie algebra is still of fundamental significance for special relativity since it is (real) isomorphic to the Lie algebra $\mathfrak{so}(3, 1)$ of the Lorentz group. To exhibit this isomorphism, one needs to first regard $\mathbf{C}^3$ as $\mathbf{R}^3 \times \mathbf{R}^3 = \mathbf{R}^6$ by way of the association of $v^i + iw^i$ with $(v^i, w^i)$. If $\{J_i, K_i, i = 1, 2, 3\}$ is the standard basis for $\mathfrak{so}(3, 1)$ that is defined by the three elementary infinitesimal Euclidian rotations and the three elementary boosts, respectively, then the isomorphism of $\mathfrak{so}(3, 1)$ with $\mathfrak{so}(3; \mathbf{C})$ as real vector spaces is defined simply by associating the basis vectors $J_i$ with the vectors $Z_i = (\mathbf{e}_i, 0)$ and the basis vectors $K_i$ with the vectors $iZ_i = (0, \mathbf{e}_i)$. More concisely, one can say that the infinitesimal Lorentz transformation $v^i J_i + w^i K_i$ goes to $(v^i + iw^i) Z_i$. By comparing the structure constants for $\mathfrak{so}(3, 1)$ in this basis:

$$[J_i, J_j] = \varepsilon_{ijk} J_k, \qquad [J_i, K_j] = \varepsilon_{ijk} K_k, \qquad [K_i, K_j] = -\varepsilon_{ijk} J_k, \tag{2.4}$$

with those of $\mathfrak{so}(3; \mathbf{R})$ relative to the basis $\{Z_i, iZ_i\}$:

$$[Z_i, Z_j] = \varepsilon_{ijk} Z_k, \qquad [Z_i, iZ_j] = i\varepsilon_{ijk} Z_k, \qquad [iZ_i, iZ_j] = -\varepsilon_{ijk} Z_k, \tag{2.5}$$

one sees that this vector space isomorphism is also a Lie algebra isomorphism.

Under this association, infinitesimal Euclidian rotations in $\mathfrak{so}(3, 1)$ become infinitesimal rotations about real axes in $\mathfrak{so}(3; \mathbf{C})$ and infinitesimal boosts become infinitesimal rotations about imaginary axes. This way of looking at Lorentz transformations as complexified rotations has its roots in the more elementary association of a real variable $x$ with the imaginary variable $ix$, which takes $e^x$ to $e^{ix}$, and consequently $\cosh x$ to $\cos x$ and $\sinh x$ to $i \sin x$. However, rather than think of Minkowksi space itself as being complexified, as when one introduces an imaginary time, it is better to think of the complex orthogonal space $\mathbf{C}^3$ with the Euclidian scalar product:

$$\gamma = \delta_{ij} Z^i \otimes Z^j \tag{2.6}$$

as simply being the new arena of relativistic geometry that replaces Minkowski space. In this expression, we have defined $\{Z^i, i = 1, 2, 3\}$ to be the coframe that is reciprocal to the frame $Z_i$, so $Z^i(Z_j) = \delta^i_j$.

The reason that $\gamma$ has the Kronecker delta for its components is because the standard basis $Z_i$ is *complex orthonormal*. When $\{Y_i, i = 1, 2, 3\}$ is not necessarily a complex orthonormal frame on $\mathbf{C}^3$ we given the tensor $\gamma$ the more general components $\gamma_{ij}$:

$$\gamma = \gamma_{ij} Y^i \otimes Y^j . \tag{2.7}$$

We denote the Lie algebra isomorphism by $\sigma: \mathfrak{so}(3, 1) \to \mathfrak{so}(3; \mathbf{C})$ and the components of the isomorphism relative to a choice of basis for both by $\sigma^{i\mu}_{j\nu}$, such that if $\omega^\mu_\nu \in \mathfrak{so}(3, 1)$ then the corresponding $\sigma^i_j \in \mathfrak{so}(3, \mathbf{C})$ is obtained by way of:

$$\sigma^i_j = \sigma^{i\mu}_{j\nu} \omega^\nu_\mu . \tag{2.8}$$

In order to make the somewhat abstract space ($\mathbf{C}^3$, $\gamma$) more physically meaningful, one needs to show how it relates to the vector spaces of bivectors and 2-forms on Minkowski space.

## 3 The geometry of bivectors

In order to make the shift from starting with the tangent bundle $T(M)$ given a Lorentzian metric $g$ to starting with the bundle of 2-forms $\Lambda^2(M)$ given an almost-complex structure *, we must first look at how everything is defined in terms of the vector space $A_2(\mathbf{R}^4) = \mathbf{R}^4 \wedge \mathbf{R}^4$ of bivectors on $\mathbf{R}^4$, and which is the exterior product of $\mathbf{R}^4$ with itself, and its dual vector space $A^2(\mathbf{R}^4) = \mathbf{R}^{4*} \wedge \mathbf{R}^{4*}$ of (algebraic) 2-forms, which is the exterior product of the dual of $\mathbf{R}^4$ with itself. For brevity, we shall simply refer to the two vector spaces as $A_2$ and $A^2$, respectively. If it should be necessary to refer to the exterior tensor products of other vector spaces, we shall modify the notation accordingly.

### 3.1 Real vector space structure on $A_2$

Since $A_2$ is six-dimensional as a real vector space, any frame – i.e., basis – for it must consist of six linearly independent bivectors $\{\mathbf{b}_I, I = 1, \ldots, 6\}$. Any other general linear frame $\{\mathbf{b}'_I, I = 1, \ldots, 6\}$ on $A_2$ will then be related to the former frame by an invertible linear transformation of $A_2$:

$$\mathbf{b}'_I = A_I^J \mathbf{b}_J \qquad (3.1)$$

that can be defined by the invertible 6×6 real matrix $A_I^J$ of components with respect to the frame $\mathbf{b}_I$. Hence, the general linear 6-frames on $A_2$ are in one-to-one correspondence with the elements of the group $GL(6; \mathbf{R})$.

A particularly useful type of frame on $A_2$ is obtained by starting with a 4-frame $\{\mathbf{e}_\mu, \mu = 0, \ldots, 3\}$ on $\mathbf{R}^4$ and considering the independent bivectors defined by all of the $\mathbf{e}_\mu \wedge \mathbf{e}_\nu$. In particular, we denote them by:

$$\mathbf{b}_i = \mathbf{e}_0 \wedge \mathbf{e}_i, \ i = 1, 2, 3, \qquad \mathbf{b}_4 = \mathbf{e}_2 \wedge \mathbf{e}_3, \quad \mathbf{b}_5 = \mathbf{e}_3 \wedge \mathbf{e}_1, \quad \mathbf{b}_6 = \mathbf{e}_1 \wedge \mathbf{e}_2. \qquad (3.2)$$

The last three vectors can be summarized in the formula:

$$\mathbf{b}_{i+3} = \tfrac{1}{2} \varepsilon_i^{jk} \mathbf{e}_j \wedge \mathbf{e}_k, \qquad (3.3)$$

in which we have lowered the first index of the Levi-Civita $\varepsilon$ symbol by means of the Kronecker delta $\delta_{ij}$. Note that this expression is not invariant under a general linear transformation of $\mathbf{R}^4$, but only under a linear transformation of the three-dimensional subspace $\Pi_3$ of $\mathbf{R}^4$ that is spanned by the 3-frame $\{\mathbf{e}_i, i = 1, 2, 3\}$ that preserves the volume element that is defined on it by:

$$V_3 = \theta^1 \wedge \theta^2 \wedge \theta^3 = \frac{1}{3!} \varepsilon_{ijk} \theta^i \wedge \theta^j \wedge \theta^k, \qquad (3.4)$$

as well as the Euclidian scalar product that is defined by:

$$\delta = \delta_{ij} \theta^i \otimes \theta^j, \qquad (3.5)$$

in which $\{\theta^i, i = 1, 2, 3\}$ is the 3-frame that is reciprocal to the $\mathbf{e}_i$. Hence, the 3-frame $\mathbf{b}_{i+3}$ is really an $SO(3; \mathbf{R})$-invariant construction. However, under complexification this will not be as much of a restriction in the eyes of the Lorentz group.

Another observation that must be made at this point is that not every frame on $A_2$ is representable in the form (3.2) for some corresponding frame on $\mathbf{R}^4$. For one thing, the set of linear frames on $\mathbf{R}^4$ is parameterized by the sixteen-dimensional Lie group $GL(4; \mathbf{R})$, while the set of linear frames on $A_2$ is parameterized by the 36-dimensional Lie group $GL(6; \mathbf{R})$. For another, not all frames on $A_2$ can be subdivided into a set of three elements with a common exterior factor – such as $\mathbf{e}_0$ – and another three elements that represent the vector space $A_2(\mathbf{R}^4)$; the other possibility is that no three of the frame members will have a common exterior factor.

The dual space to $A_2$ is the space $A^2$ of 2-forms on $\mathbf{R}^4$. Hence, a 2-form $\alpha$ can be regarded as a linear functional on the space of bivectors. In fact, since $\alpha$ is also a *bilinear* functional on $\mathbf{R}^4$ to begin with we can express the evaluation of the linear function $\alpha$ on a decomposable bivector $\mathbf{c} = \mathbf{a} \wedge \mathbf{b}$ by way of:

$$\alpha(\mathbf{c}) = \alpha(\mathbf{a}, \mathbf{b}) . \tag{3.6}$$

One then extends to non-decomposable bivectors by linearity.

Any frame $\mathbf{b}_I$ for $A_2$ defines a reciprocal 6-coframe $b^I$ on $A^2$ by way of:

$$b^I(\mathbf{b}_J) = \delta^I_J . \tag{3.7}$$

One verifies that if the reciprocal 4-coframe to $\mathbf{e}_\mu$ on $\mathbf{R}^4$ is $\theta^\mu$ then the reciprocal coframe $b^I$ to the frame $\mathbf{b}_I$ that was defined by (3.2) is given by:

$$b^i = \theta^0 \wedge \theta^i, \; i = 1, 2, 3, \qquad b^4 = \theta^2 \wedge \theta^3, \quad b^5 = \theta^3 \wedge \theta^1, \quad b^6 = \theta^1 \wedge \theta^2. \tag{3.8}$$

If one has chosen a unit-volume element for $\mathbf{R}^4$, that is, a non-vanishing 4-form:

$$V = \theta^0 \wedge \theta^1 \wedge \theta^2 \wedge \theta^3 = \frac{1}{4!} \varepsilon_{\kappa\lambda\mu\nu} \theta^\kappa \wedge \theta^\lambda \wedge \theta^\mu \wedge \theta^\nu, \tag{3.9}$$

then one can use it define a scalar product on $A_2$ by way of:

$$\langle \mathbf{F}, \mathbf{G} \rangle = V(\mathbf{F} \wedge \mathbf{G}); \tag{3.10}$$

If $\mathbf{F} = \tfrac{1}{2} F^{\mu\nu} \mathbf{e}_\mu \wedge \mathbf{e}_\nu$ and $\mathbf{G} = \tfrac{1}{2} G^{\mu\nu} \mathbf{e}_\mu \wedge \mathbf{e}_\nu$ then:

$$\langle \mathbf{F}, \mathbf{G} \rangle = \varepsilon_{\kappa\lambda\mu\nu} F^{\kappa\lambda} G^{\mu\nu}. \tag{3.11}$$

One can also express $\mathbf{F}$ and $\mathbf{G}$ as $F^I \mathbf{b}_I$ and $G^I \mathbf{b}_I$, in which case:

$$\langle \mathbf{F}, \mathbf{G} \rangle = \varepsilon_{IJ} F^I G^J, \tag{3.12}$$

in which the 6×6 real matrix $\varepsilon_{IJ}$ is equal to:

$$\varepsilon_{IJ} = \langle \mathbf{b}_I, \mathbf{b}_J \rangle = \begin{bmatrix} 0 & I \\ \hline I & 0 \end{bmatrix}. \tag{3.13}$$

By the change of frame, this can be brought into the canonical form:

$$\varepsilon'_{IJ} = \begin{bmatrix} I & 0 \\ \hline 0 & -I \end{bmatrix}, \tag{3.14}$$

so this scalar product is of signature type (+1, +1, +1, −1, −1, −1).

The quadric hypersurface in $A_2$ that is defined by the equation:

$$0 = <\mathbf{F}, \mathbf{F}> = \varepsilon_{IJ} F^I F^J, \qquad (3.15)$$

which is called the *Klein quadric*, has considerable significance in the eyes of projective geometry. A bivector **F** satisfies (3.15) iff it is decomposable; viz., it is of the form $\mathbf{a} \wedge \mathbf{b}$ for some vectors $\mathbf{a}, \mathbf{b} \in \mathbf{R}^4$. Furthermore, $\mathbf{F} \neq 0$ iff **a** and **b** are not collinear. Hence, in that case, they span a 2-plane $\Pi_2$ in $\mathbf{R}^4$. Had one chosen another pair of non-collinear vectors **c** and **d** to span $\Pi_2$ then there would be an invertible 2×2 real matrix $a_{ij}$ such that:

$$\mathbf{c} = a_{11}\mathbf{a} + a_{12}\mathbf{b}, \qquad (3.16a)$$
$$\mathbf{d} = a_{21}\mathbf{a} + a_{22}\mathbf{b}, \qquad (3.16b)$$

in which case:

$$\mathbf{c} \wedge \mathbf{d} = \det(a)\, \mathbf{a} \wedge \mathbf{b}. \qquad (3.17)$$

Hence, there is a one-to-one correspondence between 2-planes in $\mathbf{R}^4$ and equivalence classes of decomposable bivectors that differ by a non-zero scalar multiple. One can think of this as defining an embedding of the Grassmanian manifold $G_{2,4}$ of 2-planes in $\mathbf{R}^4$ into the projective space defined by all lines through the origin of $A_2$, and which is commonly referred to as the *Plücker-Klein embedding*.

One can represent a 3-plane $\Pi_3$ in $A_2$ in two ways: As long as one also has a line L through the origin of $\mathbf{R}^4$ that does not lie in $\Pi_3$, one can choose any vector **t** that generates L and define the linear injection $e_\mathbf{t}: \Pi_3 \to A_2$, $\mathbf{v} \mapsto \mathbf{t} \wedge \mathbf{v}$. A different choice of **t** would change the value of $e_\mathbf{t}(\mathbf{v})$ by only a non-zero scalar multiple. Hence, the 3-plane $\Pi_3$ gets mapped to a unique 3-dimensional subspace of $A_2$. Secondly, the exterior algebra $A_2(\Pi_3)$ of bivectors over $\Pi_3$ defines another 3-dimensional subspace of $A_2$, which is, moreover, complementary to the latter one. As shown in [**6**], these are the only two possible types of 3-dimensional subspaces in $A_2$. As discussed in [**7**], this relationship between 1+3 decompositions of $\mathbf{R}^4$ and 3+3 decompositions of $A_2$ has significance in both projective geometry and special relativity when one regards $\Pi_3$ as the rest space for a measurement.

Of course, dual statements to all of the foregoing can be made for the vector space $A^2$. In particular, one defines the scalar product on $A^2$ by means of the volume element **V** on $T^*(M)$ that is reciprocal to $V$, so $V(\mathbf{V}) = 1$:

$$<F, G> = (F \wedge G)(\mathbf{V}). \qquad (3.18)$$

If one decomposes $F$ and $G$ into $E_i b^i + B_i b^{i+3}$ and $E'_i b^i + B'_i b^{i+3}$ then

$$<F, G> = \varepsilon^{\kappa\lambda\mu\nu} F_{\kappa\lambda} G_{\mu\nu} = \varepsilon^{IJ} F_I G_J = \mathbf{E} \cdot \mathbf{B}' + \mathbf{E}' \cdot \mathbf{B}, \qquad (3.19)$$

in which the dot signifies the Euclidian scalar product on $\mathbf{R}^3$. In particular, one has:

$$<F, F> = \varepsilon^{\kappa\lambda\mu\nu} F_{\kappa\lambda} F_{\mu\nu} = \varepsilon^{IJ} F_I F_J = 2\mathbf{E} \cdot \mathbf{B}, \qquad (3.20)$$

an expression that has considerable significance in the theory of electromagnetism.

### 3.2 Complex vector space structure on $A_2$

Now let us assume that $A_2$ is endowed with a complex structure, in the form of a linear isomorphism $*: A_2 \to A_2$ that has the property $*^2 = -I$. When $\mathbf{R}^4$ is given the Lorentzian structure that makes it into Minkowski space and the unit-volume element, as above, such an isomorphism can be defined by Hodge duality.

One can think of Hodge duality as obtained from the composition of two isomorphisms: *Poincaré duality*, which comes from the unit-volume element, and takes the form:

$$\#: A_2 \to A^2, \quad \mathbf{a} \wedge \mathbf{b} \mapsto i_{\mathbf{a} \wedge \mathbf{b}} V, \tag{3.21}$$

and the metric isomorphism:

$$g^{-1} \wedge g^{-1}: A^2 \to A_2, \quad \alpha \wedge \beta \mapsto g^{-1}(\alpha) \wedge g^{-1}(\beta) \tag{3.22}$$

that represents the "raising of the indices" operation on the components of the 2-form $\alpha \wedge \beta$.

In pre-metric electromagnetism [**8, 9**], the isomorphism (3.21) is replaced with a linear electromagnetic constitutive law. Although this has deep and far-reaching physical ramifications, and defines the motivation for the present study, nevertheless, we shall make only occasional mention of this electromagnetic aspect of doing geometry in terms of the bundle of 2-forms instead of the tangent bundle, since our present concern is gravitation.

One notes that the basis (3.2) for $A_2$ that we have been using heretofore has the property that under Hodge duality one has:

$$*\mathbf{b}_i = \mathbf{b}_{i+3}, \qquad *\mathbf{b}_{i+3} = -\mathbf{b}_i. \tag{3.23}$$

Hence, if we define $A_2^{Re}$ to be the three-dimensional subspace of $A_2$ that is spanned by the $\mathbf{b}_i$, $i = 1, 2, 3$ and $A_2^{Im}$ to be the subspace spanned by the $\mathbf{b}_{i+3}$, $i = 1, 2, 3$ then since these subspaces are complementary in $A_2$ we can decompose $A_2$ into the direct sum $A_2^{Re} \oplus A_2^{Im}$. Since (3.23) implies that $*A_2^{Re} = A_2^{Im}$ and $*A_2^{Im} = A_2^{Re}$, we suspect that we are indeed justified in thinking of the subspaces $A_2^{Re}$ and $A_2^{Im}$ as the "real" and "imaginary" subspaces of $A_2$ when given the complex structure $*$.

In order to confirm this, we show how $*$ allows us to define a complex vector space structure on $A_2$ and a (non-canonical) isomorphism of $A_2$ with $\mathbf{C}^3$. In order to accomplish the first objective, it is sufficient to show how $*$ allows us to define complex scalar multiplication on $A_2$. This follows from the basic definition:

$$i\mathbf{F} = *\mathbf{F}, \tag{3.24}$$

which one then extends to any complex scalar $\alpha + i\beta$ by way of:

$$(\alpha + i\beta)\mathbf{F} = \alpha\mathbf{F} + \beta *\mathbf{F} . \tag{3.25}$$

We can then express any bivector $\mathbf{F} = E^i \mathbf{b}_i + B^i *\mathbf{b}_i$ as:

$$\mathbf{F} = (E^i + i B^i) \mathbf{b}_i . \tag{3.26}$$

Hence, the three linearly independent bivectors $\{\mathbf{b}_i, i= 1, 2, 3\}$ define a complex basis for the complex vector space $A_2$ with the complex structure *, and an isomorphism with $\mathbf{C}^3$ can be defined by any such choice of complex 3-frame:

$$\mathbf{b}: \mathbf{C}^3 \to A_2, \quad (z^1, z^2, z^3) \mapsto z^i \mathbf{b}_i . \tag{3.27}$$

One then sees that not every real 6-frame on $A_2$ will define a complex 3-frame, but only the ones for which (3.23) is satisfied. Similarly, two complex 3-frames on $A_2$ will differ by an element of $GL(3; \mathbf{C})$, which can be represented in $GL(6; \mathbf{R})$ by means of invertible 6×6 real matrices that commute with *, whose matrix with respect to the $\mathbf{b}_I$ frame is:

$$[*]_J^I = \begin{bmatrix} 0 & I \\ \hline -I & 0 \end{bmatrix}. \tag{3.28}$$

The real 6×6 matrices that represent invertible complex transformations then take the form:

$$\begin{bmatrix} A & B \\ \hline -B & A \end{bmatrix} = \begin{bmatrix} A & 0 \\ \hline 0 & A \end{bmatrix} + * \begin{bmatrix} B & 0 \\ \hline 0 & B \end{bmatrix}. \tag{3.29}$$

The association of a 3×3 complex matrix of the form $A + iB$ with its representative in $GL(6; \mathbf{R})$ then becomes clear. This representation of $GL(3; \mathbf{C})$ in $GL(6; \mathbf{R})$ is, moreover, faithful; i.e., it is one-to-one.

One must observe that the concept of a line through the origin of $A_2$ takes on a different meaning when one uses complex scalars instead of real ones. In particular, just as the complex line $\mathbb{C}$ is two-dimensional as a real vector space, similarly, a complex line through the origin of $A_2$ is a real 2-plane in $A_2$. If one expresses an arbitrary complex scalar $\alpha$ in polar form as $\alpha = ae^{i\theta}$ then one sees that such a 2-plane is obtained by starting with a bivector $\mathbf{F}$, extending to the real line [$\mathbf{F}$] that it generates, and rotating that line around a unique plane that is associated with $\mathbf{F}$ that we shall call the *duality plane;* similarly, the rotation in question will be called a *duality rotation.* The duality plane through a bivector $\mathbf{F}$ is spanned by $\mathbf{F}$ and *$\mathbf{F}$.

If one represents $\mathbf{F}$ as $E^i \mathbf{b}_i + B^i *\mathbf{b}_i$ then if we express the complex scalar $\alpha$ as $a \cos \theta + i\, a \sin \theta$ the product of $\alpha$ with $\mathbf{F}$ takes the real form:

$$\alpha F = a(E^i \cos\theta - B^i \sin\theta)\mathbf{b}_i + a(E^i \sin\theta + B^i \cos\theta)*\mathbf{b}_i. \tag{3.30}$$

One can clearly see that if the **E** and **B** vectors define a plane in $\mathbf{R}^4$ – which one calls the *polarization plane* – then the effect of a duality rotation on them is a rotation of both of them in the polarization plane through the same angle as the duality rotation in $A_2$.

If one represents Maxwell's sourceless equations for a 2-form $F$ as:

$$dF = d*F = 0 \tag{3.31}$$

then it is clear that if $F$ is a solution to these equations then so is $\alpha F$. Hence, one suspects that Maxwell's equations are more intrinsically formulated in complex terms than in real terms since duality rotations are already a symmetry of the solutions, even in the real case.

By any choice of complex 3-frame, the space $\mathbf{CPA}_2$ of all complex lines through the origin – i.e., duality planes – in $A_2$ is shown to be diffeomorphic to the complex projective space $\mathbf{CP}^2$. One must be careful not to confuse $\mathbf{CPA}_2$ with the space $\mathbf{RPA}_2$ of all real lines through the origin, which is then diffeomorphic to $\mathbf{RP}^5$.

If one assumes that the complex structure * is *self-adjoint*, so:

$$\langle \mathbf{F}, *\mathbf{G}\rangle = \langle *\mathbf{F}, \mathbf{G}\rangle, \tag{3.32}$$

then one can use * to define another real scalar product on $A_2$:

$$(\mathbf{F}, \mathbf{G}) = \langle \mathbf{F}, *\mathbf{G}\rangle = V(\mathbf{F} \wedge *\mathbf{G}) = *_{IJ} F^I G^J. \tag{3.33}$$

Assuming the self-adjointness of * is equivalent to assuming the symmetry of $*_{IJ}$.

One finds that the matrix of this scalar product with respect to the $\mathbf{b}_I$ frame is:

$$*_{IJ} = \begin{bmatrix} I & 0 \\ 0 & -I \end{bmatrix}. \tag{3.34}$$

This is already in canonical form, so we see that this scalar product has the same signature type as $\langle .\, \rangle$; indeed, they differ only by an imaginary rotation.

The **E**-**B** form of this scalar product is then:

$$(\mathbf{F}, \mathbf{G}) = \mathbf{E} \cdot \mathbf{E}' - \mathbf{B} \cdot \mathbf{B}', \tag{3.35}$$

which includes the special case:

$$(\mathbf{F}, \mathbf{F}) = E^2 - B^2. \tag{3.36}$$

This expression also has considerable significance in electromagnetism. Indeed, a necessary – but not sufficient – condition for a 2-form $F$ to represent a wavelike solution of the Maxwell equations is that both $\langle F, F\rangle$ and $(F, F)$ must vanish. A counter-example to the converse is the example of two static electric and magnetic fields that are perpendicular and have the same field strength.

When $\langle \mathbf{F}, \mathbf{F} \rangle$ and $\langle \mathbf{F}, *\mathbf{F} \rangle$ both vanish for a bivector $\mathbf{F}$ one calls it *isotropic*. Both conditions define quadric hypersurfaces in $A_2$ so the isotropic bivectors define the intersection of the two quadrics, which will then be four-dimensional as a real algebraic variety. Since both conditions are also algebraically homogeneous they also define quadrics on $\mathbf{R}PA_2$, and their intersection is represented as a three-dimensional real projective variety in it.

As we have observed above, when $\langle \mathbf{F}, \mathbf{F} \rangle = 0$ the bivector $\mathbf{F}$ is associated with a unique 2-plane $\Pi_2$ in $\mathbf{R}^4$. Since the isomorphism $*$ takes decomposable bivectors to other decomposable bivectors – i.e., it leaves the Klein quadric invariant – one also has a unique 2-plane $\Pi'_2$ in $\mathbf{R}^4$ that is associated with the bivector $*\mathbf{F}$. What the expression $(\mathbf{F}, \mathbf{F}) = \langle \mathbf{F}, *\mathbf{F} \rangle$ then tells us, by its vanishing or not, is whether the 2-planes $\Pi_2$ and $\Pi'_2$ intersect in more than the origin or not, respectively. Moreover, since the two 2-planes cannot be identical, their intersection must be either the origin or a unique line through the origin of $\mathbf{R}^4$. Indeed, one can define this line to be *lightlike*, and the resulting hypersurface in $\mathbf{R}^4$ can be shown to be a cone generated by a sphere; i.e., a light cone. This is a particularly elegant geometric way of showing that a complex structure on $A_2$ induces a conformal Lorentzian structure on $\mathbf{R}^4$.

We can combine the two real scalar products into a single complex Euclidian scalar product. Indeed, since there is only one possible signature type for a complex orthogonal structure on $\mathbf{C}^3$, of the two obvious ways of combining them, either choice produces an isometric complex orthogonal structure. If we choose to define:

$$\langle \mathbf{F}, \mathbf{G} \rangle_{\mathbb{C}} = (\mathbf{F}, \mathbf{G}) + i\langle \mathbf{F}, \mathbf{G} \rangle \tag{3.37}$$

then we see that the complex 3-frame $\{\mathbf{b}_i, i = 1, 2, 3\}$ is also *orthonormal*:

$$(\mathbf{b}_i, \mathbf{b}_j) = \delta_{ij}. \tag{3.38}$$

This can also be interpreted as meaning that the complex linear isomorphism that takes each triple $(z^1, z^2, z^3)$ of complex numbers to the bivector $z^i\mathbf{b}_i$ is also an isometry of the two complex orthogonal spaces; of course, we are giving $\mathbf{C}^3$ the complex Euclidian scalar product whose components with respect to the standard basis are $\delta_{ij}$.

Any other complex orthogonal 2-frame differs from the $\mathbf{b}_i$ frame by a complex orthogonal transformation, which can be represented by an invertible complex 3×3 matrix $A$ with the property that $A^{-1} = A^{\mathrm{T}}$. Hence, we have a (right) action of $O(3; \mathbf{C})$ on the manifold of complex orthogonal 3-frames on $A_2$. In order to maintain the frame invariance of a bivector $\mathbf{F} = F^i\mathbf{b}_i$ under this action, $O(3; \mathbf{C})$ must act on the components $F^i$ on the left by way of the inverse of $A$.

If one gives $A_2$ a volume element then one can speak of unit-volume – or simply *oriented* – complex orthogonal frames on $A_2$ and the frame transformations reduce to the matrices of $SO(3; \mathbf{C})$. Such a volume element can be defined by a non-zero 3-form on $A_2$. However, one must be careful not to confuse the exterior product of elements in the vector space $A_2$ with the exterior product of vectors in $\mathbf{R}^4$, which are traceable to two different tensor products over two different vector spaces; we shall use the notation $\perp$ for the exterior product over the vector space $\mathbf{C}^3$. One can define the volume element by means of the $Z^i$ coframe members as:

$$V_{\mathbb{C}} = Z^1 \perp Z^2 \perp Z^3 = \frac{1}{3!} \varepsilon_{ijk} Z^i \perp Z^j \perp Z^k. \tag{3.39}$$

We have thus finally arrived at a representation of $SO(3; \mathbf{C})$, which we have seen to be isomorphic to $SO_0(3, 1)$, on the three-dimensional complex orthogonal space $A_2$.

Just as the real and imaginary parts of the complex scalar product (3.37) define real quadrics in $A_2$ and its real projectivization $\mathbf{R}PA_2 \cong \mathbf{R}P^5$, so does the complex scalar product define a complex quadric in $A_2$ and its complex projectivization $\mathbf{C}PA_2 \cong \mathbf{C}P_2$. One sees immediately that the homogeneous complex quadratic equation:

$$\langle \mathbf{F}, \mathbf{F} \rangle_{\mathbb{C}} = \gamma_{ij} F^i F^j = 0 \tag{3.40}$$

is equivalent to the pair of real equations:

$$\langle \mathbf{F}, \mathbf{F} \rangle = (\mathbf{F}, \mathbf{F}) = 0. \tag{3.41}$$

Bivectors that belong to this quadric are called *isotropic*. One can also characterize them as decomposable bivectors such that $(\mathbf{F}, \mathbf{F})$ vanishes, which is equivalent to the geometric condition that their 2-planes intersect their dual 2-planes in a line.

In order represent $SO(3; \mathbf{C})$ on the dual vector space $A^2$, the only modification of the foregoing that is necessary is to note that the action of $3 \times 3$ complex matrices on coframes is the inverse transpose of the action on frames. That is, if $\mathbf{b}_i$ goes to $A_i^j \mathbf{b}_j$ then the reciprocal coframe $b^i$ must go to $\tilde{A}^i_j b^j$. In other words, the action of a group on the dual of vector space is contragredient to its action on that vector space.

### 3.3 Relationship to the formalism of self-dual bivectors

Since the methods of complex relativity have long since been established in the literature, it is necessary to explain how the representation of $SO(3; \mathbf{C})$ that we have described above relates to the established representation in terms of "self-dual" bivectors and 2-forms.

The essential point of departure is whether you wish to regard the statement:

$$*\mathbf{F} = \pm i\mathbf{F} \tag{3.42}$$

as two possible definitions of how the imaginary $i$ acts on the real vector space $A_2$ or as an eigenvalue equation for the operator *. The latter interpretation is the traditional one.

In order for (3.42) to make mathematical sense, $\mathbf{F}$ must belong to a complex vector space. However, instead of simply using the complex structure that has been given to $A_2$ by means of *, it has been traditional all along to first complexify $A_2$ and then decompose the complexified vector space, which is *six*-dimensional as a complex vector space, into two eigenspaces of *, which will both be three-dimensional as complex vector spaces. The elements of the positive eigenspace are referred to as *self-dual* bivectors and the elements of the negative eigenspace are called *anti-self-dual* bivectors. One then represents $SO(3; \mathbf{C})$ by way of self-dual complex 3-frames.

This aforementioned program seems conceptually understandable, since it represents an attempt to duplicate the corresponding constructions that were ventually established by gauge field theorists (see Atiyah, Hitchin, Singer [**10**]) for Riemannian manifolds, in which the eigenvalues of * will be real and the eigenspaces will exist in $A_2$ without complexification. However, if one approaches the basic problem of constructing a representation space for $SO(3; \mathbf{C})$ in terms of bivectors and 2-forms objectively, it also becomes clear that there is something redundant about complexifying $A_2$ when one has already given it a complex structure, especially since one only uses "half" of the resulting space.

Another point that needs to be made is how the real basis $\mathbf{b}_i$, $*\mathbf{b}_i$, $i = 1, 2, 3$ relates the "complex conjugate" bases $\mathbf{Z}_i$, $\overline{\mathbf{Z}}_i$, $i = 1, 2, 3$ that are commonly used in complex relativity, as well as in complex manifolds, in general. Although the space $A_2$ of bivectors has been given a complex structure, in order to define the conjugation of a bivector one must also assume that $A_2$ has been given a specific choice of "real + imaginary" decomposition $A_2 = A_2^{Re} \oplus A_2^{Im}$, with $A_2^{Im} = *A_2^{Re}$, so any bivector $\mathbf{F}$ can be expressed as $\mathbf{F}_{Re} + i\mathbf{F}_{Im} = \mathbf{F}_{Re} + *\mathbf{F}_{Im}$, in which both $\mathbf{F}_{Re}$ and $\mathbf{F}_{Im}$ belong to $A_2^{Re}$. Since such a decomposition is equivalent to a choice of real 3-plane $A_2^{Re}$ in $A_2$, this is physically related, but not equivalent to a choice of 3-plane – i.e., rest space – in $\mathbf{R}^4$, although we shall not dwell on that fact here. (See Delphenich [**7**].)

One can then define the complex conjugate of the bivector $\mathbf{F}$ by way of:

$$\overline{\mathbf{F}} = \mathbf{F}_{Re} - *\mathbf{F}_{Im}. \tag{3.43}$$

One then sees that, as would be consistent with life in $\mathbf{C}^3$:

$$\mathbf{F}_{Re} = \tfrac{1}{2}(\mathbf{F} + \overline{\mathbf{F}}), \qquad \mathbf{F}_{Im} = \tfrac{1}{2}(\mathbf{F} - \overline{\mathbf{F}}). \tag{3.44}$$

Note that in order for six bivectors $\mathbf{Z}_i, \overline{\mathbf{Z}}_i$ to define a real basis for $A_2$ they must all have non-vanishing real and imaginary parts. In fact, we can define such a basis by starting with the $\mathbf{b}_i$, $*\mathbf{b}_i$ by the obvious definitions:

$$\mathbf{Z}_i = \tfrac{1}{2}(\mathbf{b}_i - *\mathbf{b}_i), \qquad \overline{\mathbf{Z}}_i = \tfrac{1}{2}(\mathbf{b}_i + *\mathbf{b}_i). \tag{3.45}$$

These bivectors as sometimes referred to as the *self-dual* and *anti-self-dual* parts of $\mathbf{b}_i$ and $*\mathbf{b}_i$, since:

$$*\mathbf{Z}_i = \overline{\mathbf{Z}}_i, \qquad *\overline{\mathbf{Z}}_i = -\mathbf{Z}_i. \tag{3.46}$$

In complex relativity, the expressions (3.45) are replaced by ones in which $i*$ appears in place of $*$, which then makes:

$$*\mathbf{Z}_i = i\mathbf{Z}_i, \qquad *\overline{\mathbf{Z}}_i = -i\overline{\mathbf{Z}}_i; \tag{3.47}$$

i.e., $\mathbf{Z}_i$ and $\overline{\mathbf{Z}}_i$ are, in fact, eigenvectors of the linear isomorphism $*$ that correspond to $\pm i$, resp.

If one normalizes the definitions (3.45) by using a factor of $1/\sqrt{2}$ in place of the factor of $1/2$ then the transformation from the $\mathbf{b}_i$, $*\mathbf{b}_i$ frame to the $\mathbf{Z}_i$, $\overline{\mathbf{Z}}_i$ frame has the matrix:

$$\frac{1}{\sqrt{2}}\begin{bmatrix} I & -I \\ I & I \end{bmatrix} = \cos\frac{\pi}{4}I - \sin\frac{\pi}{4}*, \tag{3.48}$$

which is a duality rotation through $-\pi/4$; hence, it is complex orthogonal.

Now, since we will be dealing with 2-forms that take their values in complex vector spaces – viz., $\mathbb{C}^3$ and $\mathfrak{so}(3; \mathbf{C})$ – one must note that one can make sense of the operator $i*$ for such geometric objects without needing to complexify $A^2$. One simply lets the $*$ isomorphism act on $A^2$, as usual, and then lets $i$ act on the complex vector space in which the 2-forms take their values. Hence, if $F$ is 2-form on $\mathbf{R}^4$ that takes its values in a complex vector space $V$ one defines $i*F$ as the 2-form on $\mathbf{R}^4$ with values in $V$ that takes a pair of vectors $\mathbf{v}, \mathbf{w} \in \mathbf{R}^4$ to $i*F(\mathbf{v}, \mathbf{w})$; i.e.:

$$(i*F)(\mathbf{v}, \mathbf{w}) = i(*F(\mathbf{v}, \mathbf{w})). \tag{3.49}$$

One can then identify two types of $\mathbf{R}$-linear maps $L: A^2 \to V$: the ones that commute with $*$ and $i$ and the ones that anti-commute:

$$L(*F) = \pm iL(F). \tag{3.50}$$

If one defines the operator $iL*$ to be the composition of the three maps then this takes the form:

$$iL*(F) = \mp F. \tag{3.51}$$

This allows us to make rigorous sense of the concept of self-dual and anti-self-dual 2-forms without first complexifying $A^2$.

The mathematically astute reader will immediately object that we are still implicitly complexifying $A^2$ by the fact that one can define the complexification of any real vector space $V$ to be the vector space of $\mathbf{R}$-linear maps from $V$ into $\mathbf{C}$. However, we will not need to consider 2-forms with values in $\mathbf{C}$, only ones with values in $\mathbf{C}^3$ and $\mathfrak{so}(3, \mathbf{C})$. Furthermore, in the former case, we shall be primarily concerned with 2-forms of a particular sort, ones that take real bivectors to real values and imaginary bivectors to imaginary values.

Given the definition of $i*$ as in (3.51), one can then polarize any $L$ into its commuting and anti-commuting parts in the usual way:

$$L_+ = \tfrac{1}{2}(L - i*L), \qquad L_- = \tfrac{1}{2}(L + i*L) . \tag{3.52}$$

When $A_2$, hence $A^2$, is given a complex structure *, there is a canonical real 2-form on $\mathbf{R}^4$ with values in $\mathbf{C}^3$ that is defined by the isomorphism of $A^2$, when given the standard basis defined by $b^i$, with the standard basis for $\mathbf{C}^3$. Since the $b^i$ are also 2-forms of the form $\theta^\mu \wedge \theta^\nu$, we define the $\mathbf{C}^3$-valued canonical 2-form $Z^i$ on $\mathbf{R}^4$ by:

$$Z^i = \theta^0 \wedge \theta^i + i*(\theta^0 \wedge \theta^i). \tag{3.53}$$

This 2-form then has the property it takes a real bivector, such as any $\mathbf{e}_0 \wedge \mathbf{e}_i$, to a real 3-vector and an imaginary bivector, such as the $\mathbf{e}_i \wedge \mathbf{e}_j$, to an imaginary 3-vector.

When the operator $i*$ acts on $Z^i$, one gets:

$$i*Z^i = i*(\theta^0 \wedge \theta^i) + \theta^0 \wedge \theta^i = Z^i; \tag{3.54}$$

i.e., it is self-dual. One can easily see that $\bar Z^i$ is then anti-self-dual.

We can also express the 2-form $Z^i$ in the manner that is more familiar from conventional complex relativity:

$$Z^i = \frac{1}{2} Z^i_{\mu\nu} \theta^\mu \wedge \theta^\nu = Z^i_{0j} \theta^0 \wedge \theta^j + \frac{i}{2} Z^i_{jk} \theta^j \wedge \theta^k , \tag{3.55}$$

in which:

$$Z^i_{0j} = -Z^i_{j0} = \delta^i_j, \qquad Z^i_{jk} = \varepsilon_{ijk}, \qquad \text{(all other are zero).} \tag{3.56}$$

Similarly, one can then use the inverses $Z_i^{\mu\nu}$ of the $Z^i_{\mu\nu}$ matrices – which are represented by the same matrices as in (3.56) – to map the components $F_{\mu\nu}$ of a 2-form $F$ with respect to the coframe $\theta^\mu$ to the corresponding complex components of a vector in $\mathbf{C}^3$:

$$F_i = Z_i^{\mu\nu} F_{\mu\nu} = Z_i^{0j} F_{0j} + \frac{i}{2} Z_i^{jk} F_{jk} . \tag{3.57}$$

However, one will note in the sequel that the only point at which it becomes necessary for us to introduce the operator $i*$ into our geometrical discussion will be when we wish to duplicate the Debever-Penrose decomposition of the Riemannian curvature tensor into the Weyl curvature tensor, the trace-free part of the Ricci curvature, and the scalar curvature.

## 4 Associated principal bundles for the vector bundle of 2-forms

Next, we must extend the scope of our previous discussion from vector spaces to vector bundles. In the case of the real vector spaces $A_2$ and $A^2$, we simply pass to the real vector bundles $\Lambda_2(M)$ and $\Lambda^2(M)$ that consist of all bivector fields and differential 2-forms on the spacetime manifold $M$. The fibers of these bundles will then be (non-canonically) isomorphic to the vector spaces $A_2$ and $A^2$, respectively. The isomorphism for a given fiber of $\Lambda^2_x(M)$ can be specified by a choice of frame on that fiber. However, unless $M$ is parallelizable this choice of frame cannot be made globally.

In the case of a real vector bundle, the assignment of an isomorphism * such that $*^2 = -I$ to each fiber is not referred to as a complex structure, but an *almost-complex* structure. This is because when the bundle in question is the tangent bundle to a real manifold one cannot necessarily find a complex atlas for the manifold that has the bundle in question for its complex tangent bundle. The issue is one of *integrability*, but since we are chiefly concerned with a real vector bundle whose fiber dimension is not equal to the dimension of the tangent spaces to the base manifold, integrability cannot be an issue, anyway.

We need to discuss the various ways that one can associate a principle fiber bundle with the vector bundles $\Lambda_2(M)$ and $\Lambda^2(M)$. Basically, this amounts to looking at the possible types of frames that one can choose in the fibers.

Suppose $G$ is a Lie group and one has defined a representation of $G$ in an $n$-dimensional (real or complex) vector space $V$; i.e., a linear action of $G$ on $V$. A $G$-principal bundle $P \to M$ can then be associated with a vector bundle $E \to M$, whose typical fiber is isomorphic to $V$. Conversely, in order to define an associated principal bundle for $E$ one must first settle on what group is acting on the bundle. Equivalently, one identifies a class of frames in the fibers. Although a frame in a fiber $E_x(M)$ can simply be regarded as a maximal set of linearly independent vectors $\{\mathbf{e}_i, i = 1, \ldots, n\}$, it is often more convenient to think of a frame as a linear isomorphism $\mathbf{e}: \mathbb{K}^n \to E_x(M)$ that takes the canonical basis for $\mathbb{K}^n$ to that set of vectors, in which we let $\mathbb{K}$ represent either $\mathbf{R}$ or $\mathbf{C}$. A coframe then becomes a linear isomorphism $\theta: E_x(M) \to \mathbf{K}^n$; i.e., a 1-form with values in $\mathbf{K}^n$. In particular, the *reciprocal* coframe to the frame $\mathbf{e}$ is the one that makes $\theta(\mathbf{e}) = I$, or, in terms of individual frame members:

$$\theta^i(\mathbf{e}_j) = \delta^i_j. \tag{4.1}$$

In general, one can always consider all of the linear frames in a fiber, which then defines an associated $GL(n)$-principal bundle, where we intend this to mean $GL(n; \mathbf{R})$ or $GL(n; \mathbf{C})$ depending upon whether $V$ is real or complex, respectively. If the group $G$ is a subgroup of $GL(n)$ then one restricts oneself to some set of frames that all differ by the elements of $G$. For instance, if $G$ is $O(n)$ then one restricts oneself to a set of orthogonal frames. In general, any frame defines such a set by its orbit under the action of $G$. The associated $O(n)$-principal bundle is then defined by a choice of a set of orthogonal frames at each point; i.e., by a choice of $G$-orbit.

In the case of the bundle $\Lambda^2(M)$ of 2-forms on a four-dimensional Lorentzian manifold $M$ there are a number of groups and associated frames that one can consider. If one regards a typical fiber $\Lambda_x^2(M)$ as the six-real-dimensional vector space $\mathbf{R}^6$ then the $G$ in question would be $GL(6; \mathbf{R})$ and the associated principal bundle would be defined by all real linear frames in the fibers. However, since we are assuming that $\Lambda^2(M)$ has an almost-complex structure * defined on it, it would be more to the point to assume that the vector space is $\mathbf{C}^3$, the group is $GL(3; \mathbf{C})$, and the associated principal bundle is defined by all complex 3-frames in the fibers.

Of course, from the standpoint of relativity theory, which treats geometry as something that lives in the metric structure on the tangent bundle, the role of the bundle $\Lambda^2(M)$ is somewhat secondary, since one starts with the action of the Lorentz group $SO(3, 1)$ on tangent frames and defines its action on 2-forms by way of the tensor product representation. In particular, if $\mathbf{e}_\mu$, $\mu = 0, \ldots, 3$ is a Lorentzian frame for Minkowski space and the right action of a Lorentz transformation $A$ takes the frame $\mathbf{e}_\mu$ to $A^{-1}\mathbf{e}_\mu$ then the frame on $\Lambda^2(\mathbf{R}^4)$ that is defined by $\mathbf{e}_\mu \wedge \mathbf{e}_\nu$, $\mu < \nu$ goes to the frame $A^{-1}\mathbf{e}_\mu \wedge A^{-1}\mathbf{e}_\nu$. This gives the usual transformation of tensor components that takes $F_{\mu\nu}$ to $A_\mu^\rho A_\nu^\sigma F_{\rho\sigma}$. We then have an injective homomorphism of $SO(3, 1)$ in $GL(6; \mathbf{R})$ whose image is then isomorphic to $SO(3, 1)$.

This suggests that for the purposes of relativity and the theory of gravitation we should probably wish to concentrate on the frames in $\Lambda^2(\mathbf{R}^4)$ that make this isomorphism take the form of the isomorphism of $SO_0(3, 1)$ with $SO(3; \mathbf{C})$. These would then be the oriented complex orthogonal 3-frames, relative to the almost-complex structure and orthogonal structure that is defined by *. We can define a complex orthogonal structure $\gamma$ on $\Lambda^2(M)$ locally by the tensor field:

$$\gamma = \gamma_{ij} Z^i \otimes Z^j, \qquad (4.2)$$

whose restriction to each fiber is the complex orthogonal scalar product that we defined by means of * and $V$. The tensor field $\gamma$ also defines a map $\gamma: GL(3;\mathbf{C})(\Lambda) \to GL(3;\mathbf{C})/SO(3;\mathbf{C})$ that takes the complex frame $Z^i$ to the components $\gamma_{ij}$ of $\gamma$ with respect to that frame. A reduction of $GL(3;\mathbf{C})(\Lambda)$ to $O(3;\mathbf{C})(\Lambda)$ is then defined by the set of all complex frames that map to $\delta_{ij}$, which will then be, by definition, complex orthogonal.

Hence, by the isomorphism of the Lie groups $SO_0(3, 1)$ and $SO(3; \mathbf{C})$ the bundle of oriented complex orthogonal 3-frames in $\Lambda^2(M)$ is isomorphic to the bundle of oriented time-oriented Lorentzian frames in the tangent bundle $T(M)$. One can then relate the geometric constructions that are customarily defined relative to frames in the tangent bundle with corresponding constructions that relate to frames in the bundle of 2-forms.

We shall use the notation $SO_0(3,1)(M)$ for the bundle of oriented, time-oriented Lorentzian frames in $T(M)$ and the notation $SO(3; \mathbf{C})(\Lambda)$ for the bundle of oriented complex orthogonal 3-frames in $\Lambda^2(M)$. The isomorphism of frames can then be represented by the association of the complex orthogonal 3-frame $\mathbf{b}_i$, $i = 1, 2, 3$ with the Lorentzian 4-frame $\mathbf{e}_\mu$, $\mu = 0, \ldots, 3$, as discussed above.

The isomorphic representation of a given fiber $\Lambda_x^2(M)$ by $\mathbf{C}^3$ is then defined by a complex 3-frame in $\Lambda_x^2(M)$. Hence, the isomorphism of the fibers with $\mathbf{C}^3$ is not global in the general case of a non-parallelizable $M$, since its existence would imply the existence of a global frame field. Nevertheless, the isomorphism does exist locally over any open subset $U \subset M$ for which a local frame field exists, so we will sometimes represent a complex 3-frame field over such an open subset by either by a set of three linearly independent sections $b^i: U \to U \times \Lambda^2(\mathbf{R}^4)$, $i = 1, 2, 3$ or by a set of three linearly independent sections $Z^i: U \to U \times \mathbf{C}^3$, $i = 1, 2, 3$, and the isomorphism of $\Lambda^2(\mathbf{R}^4)$ with $\mathbf{C}^3$ will be denoted by $Z: \Lambda^2(\mathbf{R}^4) \to \mathbf{C}^3$, $F \mapsto F^i = Z^i(F)$, which makes $\{Z^i, i = 1, 2, 3\}$ a coframe on $\Lambda^2(U)$ for each $x \in U$.

The bundle $SO_0(3, 1)(M)$ has a canonical 1-form, which we also denote by $\theta^\mu$, with values in $\mathbf{R}^4$ that it inherits from the bundle $GL(M)$ of all linear frames on $M$ by restriction. The origin of the canonical 1-form on $GL(M)$ is in the fact that a linear 4-frame $\mathbf{e}_\mu$ on a tangent space $T_x(M)$ is an isomorphism of $\mathbf{R}^4$ with $T_x(M)$. Hence, the reciprocal coframe $\theta^\mu$ is an isomorphism of $T_x(M)$ with $\mathbb{R}^4$, as well as a 1-form on $M$ with values in $\mathbf{R}^4$. One can lift it to an $\mathbf{R}^4$-valued 1-form on $GL(M)$ by pulling it back along the projection $\pi: GL(M) \to M$.

One can define a canonical real 2-form $Z^i$ on $GL(3; \mathbb{C})(\Lambda)$ that takes its values in $\mathbf{C}^3$ in an analogous fashion. One starts with the fact that a complex 3-frame $Z_i$ on a fiber $\Lambda_x^2(M)$ of $\Lambda^2(M)$ defines a $\mathbf{C}$-linear isomorphism of $\mathbf{C}^3$ with $\Lambda_x^2(M)$, as discussed in the previous section. Its reciprocal coframe $Z^i$, which can be represented by the 2-forms that were defined in (3.51), then defines a $\mathbf{C}$-linear isomorphism of $\Lambda_x^2(M)$ with $\mathbf{C}^3$, as well as a real 2-form on $M$ with values in $\mathbf{C}^3$. One can then lift it to a $\mathbf{C}^3$-valued 2-form on $GL(3; \mathbf{C})(\Lambda)$ by pulling it back along the projection $GL(3; \mathbf{C})(\Lambda) \to M$.

In fact, the canonical 2-form on $GL(3; \mathbf{C})(\Lambda)$ associates each complex 3-frame of $GL(3; \mathbf{C})(\Lambda)$ with the $\mathbf{C}^3$-valued 2-form $Z^i$, as defined in (3.51). Hence, when one chooses a local $\mathbf{C}^3$-frame $Z: U \to GL(3; \mathbf{C})(\Lambda)$ the canonical 2-form on $GL(3; \mathbf{C})(\Lambda)$ pulls down to the $\mathbf{C}^3$-valued 2-form $Z^i$ on $U$ that is defined by (3.51).

The canonical 2-form $Z^i$ on $SO(3; \mathbf{C})(\Lambda)$ is then just the restriction of the canonical 2-form on $GL(3; \mathbf{C})(\Lambda)$ to complex orthogonal frames.

It is a useful coincidence that since the elements $\omega_\nu^\mu$ of the Lie algebra $\mathfrak{so}(3, 1)$, when their upper index is lowered, can define the components $\omega_{\mu\nu}$ of 2-forms if one is given a frame. Hence, one can also use the same $Z_{\mu\nu}^i$ matrices to define an $\mathbf{R}$-linear isomorphism of $\mathfrak{so}(3, 1)$ with $\mathfrak{so}(3, \mathbf{C})$ when one regards the three elementary infinitesimal boost matrices as the complex basis $Z^i$ for $\mathfrak{so}(3, \mathbf{C})$. One should then note that the $Z_{\mu\nu}^i$ matrices will then take the elementary infinitesimal rotations to the $iZ^i$, which is not how we defined the isomorphism of $\mathfrak{so}(3, 1)$ with $\mathfrak{so}(3, \mathbf{C})$ in the previous section.

Before we discuss the representation of general relativity in terms of the geometry of the bundle $\Lambda^2(M)$, we shall first try to briefly summarize the way it is represented in terms of the geometry of the bundle $T(M)$, where $M$ is the four-dimensional spacetime manifold.

## 5  Levi-Civita connection on the bundle $SO(3, 1)(M)$

As is traditional in the post-Cartan era of differential geometry (cf., [**11, 12**]), we introduce a connection on the bundle the bundle $SO(3, 1)(M)$ of oriented Lorentzian frames by starting with a linear connection on the bundle of $GL(M)$ of all linear frames and reducing to an $\mathfrak{so}(3, 1)$-connection. In the spirit of gauge field theory, we introduce the connection by way of its 1-form $\omega$, which is a 1-form on $GL(M)$ that takes its values in the Lie algebra $\mathfrak{gl}(4; \mathbf{R})$. Moreover, one requires that when one goes from a linear frame $\mathbf{e}_\mu$ in a tangent space $T_x(M)$ to another linear frame $\mathbf{f}_\mu = A_\mu^\nu \mathbf{e}_\nu$ that the values of $\omega$ transform to $\mathrm{Ad}^{-1}(A)\omega$. When $\omega$ is represented by a local matrix-valued 1-form $\omega_\nu^\mu$ on an open subset $U \subset M$, such as the domain of a coordinate chart $(U, x^\mu)$:

$$\omega_\nu^\mu = \Gamma_{\kappa\nu}^\mu dx^\kappa, \tag{5.1}$$

the requirement of $\mathrm{Ad}^{-1}$-equivariance of $\omega$ takes the form of saying that $A_\mu^\nu$ is a $GL(4; \mathbf{R})$-valued function on $U$ that effects a transition to another frame field $\mathbf{e}_\mu = A_\nu^\mu dx^\nu$ then the 1-form $\omega$ transforms like [1]:

$$\omega \mapsto A^{-1}\omega A + A^{-1}dA. \tag{5.2}$$

A more geometrically intuitive picture for the introduction of a linear connection on $M$ is to regard the value $\omega(\mathbf{v})$ that the local 1-form $\omega$ on $U$ associates to a tangent vector $\mathbf{v} \in T_x(U)$ as the infinitesimal generator of a linear transformation that takes any frame at $x \in U$ to a neighboring frame in the direction of $\mathbf{v}$ that one regards as parallel to the initial frame.

One introduces the *covariant differential* of a local frame field $\mathbf{e}: U \to GL(M)$, $x \mapsto \mathbf{e}_\mu(x)$ by way of:

$$\nabla \mathbf{e}_\mu = D\mathbf{e}_\mu - \omega_\mu^\nu \otimes \mathbf{e}_\nu; \tag{5.3}$$

in this expression, we are using the symbol $D$ to represent the differential of the map $\mathbf{e}$. One then calls the frame field $\mathbf{e}_\mu$ *parallel on U* iff $\nabla \mathbf{e}_\mu = 0$; i.e.:

$$D\mathbf{e}_\mu = \omega_\mu^\nu \otimes \mathbf{e}_\nu. \tag{5.4}$$

Given such a parallel frame field, one can speak of the parallelism of vectors, covectors, and tensors more generally by specifying that their components with respect to $\mathbf{e}_\mu$ and its reciprocal coframe must be constant.

When one has a general frame field on $U$, such as the natural frame field $\partial/\partial x^\mu$ for a coordinate chart $(U, x^\mu)$, one can extend to the covariant differential from frame fields to more general geometric objects by letting the connection 1-form act on the components of the object being differentiated.

---

[1] Naturally, we intend that the products in the right-hand side of this association are matrix products.

For instance, the covariant differential of a vector field $\mathbf{v} = v^\mu \partial/\partial x^\mu$ is given by:

$$\nabla \mathbf{v} = dv^\mu \otimes \frac{\partial}{\partial x^\mu} + v^\mu \otimes \nabla \frac{\partial}{\partial x^\mu} = \nabla v^\mu \otimes \frac{\partial}{\partial x^\mu}. \tag{5.5}$$

with:

$$\nabla v^\mu = dv^\mu + \omega^\mu_\nu v^\nu. \tag{5.6}$$

The vector field $\mathbf{v}$ is then said to be *parallel* iff $\nabla \mathbf{v} = 0$. Although this definition of $\nabla \mathbf{v}$ seems to violate the spirit of covariance, nevertheless, the whole point of introducing $\omega$ is precisely to make the expression (5.5) frame-invariant.

Generally, it is too strong a restriction to demand that a geometric object be parallel over all of an open subset, since that demands that it be parallel in every direction at every point. One then introduces the *covariant derivative* in the direction $\mathbf{w}$ by taking the interior product of $\mathbf{w}$ with the operator $\nabla$. For instance, the covariant derivative of the vector field $\mathbf{v}$ in the direction $\mathbf{w}$ is:

$$\nabla_\mathbf{w} \mathbf{v} = i_\mathbf{w} \nabla v^\mu \otimes \frac{\partial}{\partial x^\mu} = (w^n v_{\mu,\nu} + \Gamma^\mu_{\kappa\nu} w^\kappa v^\nu) \otimes \frac{\partial}{\partial x^\mu}. \tag{5.7}$$

A curve in $M$ is said to be *geodesic* iff its tangent vector field is parallel with respect to itself; i.e., is parallel translated along the curve. This gives a system of ordinary differential equation that can be expressed in a frame invariant way by:

$$\nabla_\mathbf{v} \mathbf{v} = 0 \tag{5.8}$$

or in local component form as:

$$\frac{dv^\mu}{d\tau} + \Gamma^\mu_{\kappa\nu} v^\kappa v^\nu = 0. \tag{5.9}$$

In general relativity, geodesics are assumed to represent the natural curves that are followed by matter – whether massive or not – in spacetime.

Dually, since $GL(4; \mathbf{R})$ acts on coframes by the inverse of the transformation that acts on frames the covariant differential of a coframe field is:

$$\nabla \theta^\mu = D\theta^\mu + \omega^\mu_\nu \otimes \theta^\nu, \tag{5.10}$$

so it is parallel iff:

$$D\theta^\mu = -\omega^\mu_\nu \otimes \theta^\nu. \tag{5.11}$$

One then defines the covariant differential of a 1-form $\alpha = \alpha_\mu \, dx^\mu$ by way of:

$$\nabla \alpha = d\alpha_\mu \otimes dx^\mu + \alpha_\mu \otimes \nabla(dx^\mu) = (d\alpha_\mu - \omega_\mu^\nu \alpha_\nu) \otimes dx^\mu. \tag{5.12}$$

The extension of the covariant differential to higher rank tensor fields follows from requiring it to be a linear derivation on tensor products:

$$\begin{aligned}\nabla(\mathbf{v} \otimes \mathbf{w} \otimes \ldots \otimes \alpha \otimes \beta) &= \\ &= (\nabla\mathbf{v}) \otimes \mathbf{w} \otimes \ldots \otimes \alpha \otimes \beta + \mathbf{v} \otimes \nabla(\mathbf{w}) \otimes \ldots \otimes \alpha \otimes \beta + \ldots \\ &\quad + \mathbf{v} \otimes \mathbf{w} \otimes \ldots \otimes \nabla(\alpha) \otimes \beta + \mathbf{v} \otimes \mathbf{w} \otimes \ldots \otimes \alpha \otimes \nabla(\beta)\end{aligned} \tag{5.13}$$

One can also define an *exterior covariant derivative* that acts on differential forms on $GL(M)$ that take their values in a vector space $V$ on which $GL(4; \mathbf{R})$ acts linearly. In general, one defines:

$$\nabla \alpha = d\alpha + \omega \wedge \alpha, \tag{5.14}$$

although the precise meaning of the second term on the right-hand side depends upon $V$ and the way that $GL(4; \mathbf{R})$ acts on it.

The *torsion* $\Theta^\mu$ and *curvature* $\Omega_\nu^\mu$ 2-forms for the connection 1-form $\omega$ are defined by the *Cartan structure equations:*

$$\Theta^\mu = \nabla \theta^\mu = d\theta^\mu + \omega_\nu^\mu \wedge \theta^\nu \tag{5.15a}$$
$$\Omega_\nu^\mu = \nabla \omega_\nu^\mu = d\omega_\nu^\mu + \omega_\kappa^\mu \wedge \omega_\nu^\kappa. \tag{5.15b}$$

In (5.15a), $\theta^\mu$ can refer to either the canonical $\mathbf{R}^4$-valued 1-form on $GL(M)$, or the local coframe field that it pulls down to by a choice of local frame field.

By a second exterior covariant derivation, one obtains the *Bianchi identities* from these equations:

$$\nabla \Theta^\mu = \nabla^2 \theta^\mu = \Omega_\nu^\mu \wedge \theta^\nu \tag{5.16a}$$
$$\nabla \Omega_\nu^\mu = \nabla^2 \omega_\nu^\mu = 0. \tag{5.16b}$$

When a connection has vanishing torsion one has an immediate algebraic identity that follows from the first Bianchi identity:

$$0 = \Omega_\nu^\mu \wedge \theta^\nu. \tag{5.17}$$

In order to reduce a linear connection on $GL(M)$ to an $SL(4; \mathbf{R})$ connection, one must assume that $M$ – or rather, $T(M)$ – is orientable to begin with and that one has chosen a unit-volume element $V$ on it. Locally, this takes the form of a non-vanishing 4-form:

$$V = dx^0 \wedge dx^1 \wedge dx^2 \wedge dx^3 = \frac{1}{4!} \varepsilon_{\kappa\lambda\mu\nu} dx^\kappa \wedge dx^\lambda \wedge dx^\mu \wedge dx^\nu. \tag{5.18}$$

When one transforms to a general linear coframe $\theta^\mu = A^\nu_\mu dx^\nu$ the components of $V$ go to $1/(4!) \det(A) \; \varepsilon_{\kappa\lambda\mu\nu}$. Hence, one can define $V$ globally as a function on $GL(M)$ by assigning a frame $\mathbf{e}_\mu$ to the real number $\det(\mathbf{e}_\mu)$; note that this expression is well-defined since if we had chosen any other initial frame field that differed by an invertible linear transformation $L$ from $dx^\mu$ the new determinant would be $\det(L^{-1}AL) = \det(A)$. The reduction to the bundle $SL(M)$ of unit-volume frames is defined by all linear frames that map to unity under this map.

In order for $\omega$ to be reducible to an $\mathfrak{sl}(4; \mathbf{R})$-connection it must parallel translate the volume element $V$ in any direction. Hence, the covariant differential of $V$ must vanish:

$$\begin{aligned}
0 &= \nabla V \\
&= -\frac{1}{4!} \varepsilon_{\kappa\lambda\mu\nu} [\omega^\kappa_\alpha dx^\alpha \wedge dx^\lambda \wedge dx^\mu \wedge dx^\nu + \ldots + dx^\kappa \wedge dx^\lambda \wedge dx^\mu \wedge \omega^\nu_\alpha dx^\alpha] \\
&= -\frac{1}{4!} \varepsilon_{\kappa\lambda\mu\nu} \omega^\alpha_\alpha dx^\kappa \wedge dx^\lambda \wedge dx^\mu \wedge dx^\nu,
\end{aligned} \tag{5.19}$$

which says that one must have:

$$\mathrm{Tr}(\omega) = 0 ; \tag{5.20}$$

i.e., $\omega$ must take its values in $\mathfrak{sl}(4; \mathbf{C})$.

One usually introduces a Lorentzian metric on the bundle $T(M)$ by defining a symmetric non-degenerate second rank covariant tensor field $g$ on $M$. It gets expressed locally in terms of an arbitrary coframe field $\theta^\mu$ on $U$ as:

$$g = g_{\mu\nu} \theta^\mu \otimes \theta^\nu . \tag{5.21}$$

Note that this construction also makes rigorous sense if $\theta^\mu$ is the canonical $\mathbf{R}^4$-valued 1-form on $GL(M)$.

In order to make it Lorentzian, one must require that there be a linear transformation $A$ at any point of $U$ such that $A^T g A$ takes the form $\eta_{\mu\nu} = \mathrm{diag}(+1, -1, -1, -1)$. Moreover, the local frame field $\theta^\mu$ is called *Lorentzian* – or *orthonormal* – iff one has:

$$g = \eta_{\mu\nu} \theta^\mu \otimes \theta^\nu . \tag{5.22}$$

In order to reduce an $\mathfrak{sl}(4; \mathbf{R})$-connection $\omega$ to a Lorentzian connection, it must preserve the Lorentzian metric under parallel translation. Hence, the covariant differential of the metric tensor field must vanish:

$$\nabla g_{\mu\nu} = dg_{\mu\nu} - \omega^\lambda_\mu g_{\lambda\nu} - \omega^\lambda_\nu g_{\lambda\mu} = 0, \tag{5.23}$$

For an orthogonal frame field the components of $g$ are $\eta_{\mu\nu} = \mathrm{diag}(+1, -1, -1, -1)$ and the metricity condition (5.23) takes the form:

$$\omega_{\mu\nu} + \omega_{\nu\mu} = 0 ; \tag{5.24}$$

i.e., the connection 1-form must take its values in $\mathfrak{so}(3, 1)$.

The *Levi-Civita* connection is the metric connection on $SO_0(3, 1)(M)$ that is uniquely defined by the requirement that its torsion vanish:

$$\Theta^\mu = d\theta^\mu + \omega^\mu_\nu \wedge \theta^\nu = 0. \tag{5.25}$$

For a general (i.e., *anholonomic*) local coframe field $\theta^\mu$ on $U \subset M$, one has:

$$d\theta^\mu = \tfrac{1}{2} c^\mu_{\kappa\nu} \theta^\kappa \wedge \theta^\nu , \tag{5.26}$$

for a unique set of functions $c^\mu_{\kappa\nu}$ on $U$. If we set:

$$\omega^\mu_\nu = \Gamma^\mu_{\kappa\nu} \theta^\kappa , \tag{5.27}$$

in which the functions $\Gamma^\mu_{\kappa\nu}$ are the *Ricci rotation coefficients* for an anholonomic frame and the *Christoffel symbols* for a holonomic frame (i.e., $d\theta^\mu = 0$) then (5.23) takes the local form:

$$\Gamma^\mu_{\kappa\nu} - \Gamma^\mu_{\nu\kappa} = - c^\mu_{\kappa\nu} . \tag{5.28}$$

Hence, for a holonomic frame, one recovers the usual notion that vanishing torsion is equivalent to the symmetry of the Christoffel symbols in the lower indices.

To make contact with the traditional components $R^\mu_{\kappa\lambda\nu}$ of the *Riemann curvature tensor* for the Levi-Civita connection $\omega$, one must set:

$$\Omega^\mu_\nu = \tfrac{1}{2} R^\mu_{\kappa\lambda\nu} \theta^\kappa \wedge \theta^\lambda . \tag{5.29}$$

In this form, one sees that the symmetries of the components follow naturally as properties of 2-form components, elements of $\mathfrak{so}(3; \mathbf{C})$, and the first Bianchi identity. Notice that one could just as well define (5.29) globally on $SO_0(3, 1)(M)$ by means of the canonical 1-form $\theta^\mu$.

The *Ricci curvature tensor* is obtained by contracting (5.29) with the local frame field $\mathbf{e}_\mu$ that is reciprocal to $\theta^\mu$:

$$\Omega_\nu = i_{\mathbf{e}_\mu} \Omega^\mu_\nu = R_{\lambda\nu} \theta^\lambda \qquad (R_{\lambda\nu} = R^\mu_{\mu\lambda\nu}) . \tag{5.30}$$

Hence, in this form the Ricci tensor is a 1-form with values in $\mathbf{R}^4$, although it is not necessarily a 4-coframe field on $U$ unless $\det(R_{\lambda\nu})$ is non-zero. The traditional form for the tensor is obtained from:

$$\Omega_\nu \otimes \theta^\nu = R_{\lambda\nu} \theta^\lambda \otimes \theta^\nu. \tag{5.31}$$

Note that this construction does not involve the metric.

In order to define the *scalar curvature R*, we must raise the index on $\Omega_\nu$ by means of $g^{\mu\nu}$ and then contract with $\mathbf{e}_\mu$:

$$R = i_{\mathbf{e}_\mu} \Omega^\mu = R^\mu_\mu. \tag{5.32}$$

This means that if we raise the $\nu$ index in $\Omega^\mu_\nu$ to begin with, we can also say that:

$$R = i_{\mathbf{e}_\mu \wedge \mathbf{e}_\nu} \Omega^{\mu\nu}. \tag{5.33}$$

The Hodge dual of this 0-form is a 4-form that can be expressed as:

$$*R = R\,V = \theta^\mu \wedge \theta^\nu \wedge *\Omega_{\mu\nu}, \tag{5.34}$$

which then represents the *Einstein-Hilbert Lagrangian* for the spacetime metric.

Closely related to this 4-form is the *curvature 3-form:*

$$*G_\mu = \theta^\nu \wedge *\Omega_{\mu\nu} = (R_{\mu\nu} - \tfrac{1}{2} R g_{\mu\nu}) *\theta^\nu, \tag{5.35}$$

whose components take the form of the *Einstein tensor* for $g_{\mu\nu}$, which is characterized by the divergenceless part of the Ricci curvature tensor. Its dual 1-form $G_\mu$ is then coupled to the stress-energy-momentum 1-form $T_\mu$ by way of the Einstein field equations for $g_{\mu\nu}$:

$$G_\mu = 8\pi\kappa T_\mu, \tag{5.36}$$

which are usually expressed in component form as:

$$R_{\mu\nu} - \tfrac{1}{2} R g_{\mu\nu} = 8\pi\kappa T_{\mu\nu}. \tag{5.37}$$

## 6 Levi-Civita connection on the bundle $SO(3; \mathbf{C})(\Lambda)$

Although we could use the isomorphism of the bundles $SO_0(3, 1)(M)$ and $SO(3; \mathbf{C})(\Lambda)$ to pull back all of the aforementioned machinery of general relativity from $SO(3, 1)(M)$ to $SO(3; \mathbf{C})(\Lambda)$, that would violate the spirit of the present work, which proposes that one can do general relativity on the bundle $SO(3; \mathbf{C})(\Lambda)$ independently of the way that one defines everything on $SO_0(3, 1)(M)$.

Now that we have our associated $SO(3; \mathbf{C})$-principal bundle to $\Lambda^2(M)$, in the form of the bundle $SO(3; \mathbf{C})(\Lambda)$ of oriented complex orthonormal frames in its fibers, we can speak of defining an $\mathfrak{so}(3; \mathbf{C})$ connection on that bundle. This means defining the infinitesimal parallel translation of any frame $Z^i$ in $SO(3; \mathbf{C})(\Lambda)$ at any point $x \in M$ to the neighboring points of $M$. Furthermore, since the motions that we are considering are volume-preserving complex orthogonal transformations, we expect that the infinitesimal parallel translations must come from the Lie algebra $\mathfrak{so}(3; \mathbf{C})$.

Since we are dealing with frames we can represent the elements of $\mathfrak{so}(3; \mathbf{C})$ by traceless anti-symmetric complex $3 \times 3$ matrices $\sigma^i_j$, which take the frame $Z^i$ to the infinitesimal increment $\sigma^i_j Z^j$, which is, of course, not necessarily a frame. At the same time, in order to preserve the frame-invariance of a complex 3-vector $F = F_i Z^i$ (i.e., a 2-form $F = \frac{1}{2} F_{\mu\nu} \theta^\mu \wedge \theta^\nu$) the components must transform by the transpose of the transformation. Hence, the $F_i$ must go to $-\sigma^j_i F_j$.

In order to relate the geometry that we define on oriented complex orthogonal frames back to the usual general relativistic picture one need only use the association of the oriented complex orthonormal 3-frame $Z^i$ with the oriented, time-oriented Lorentzian 4-frame $\theta^\mu$, so that the group $SO_0(3, 1)$ gets represented in $A_2$ and $A^2$. We then associate the $3 \times 3$ complex matrix $\sigma^i_k \in \mathfrak{so}(3; \mathbf{C})$ with the $4 \times 4$ real matrix $\omega^\mu_\nu \in \mathfrak{so}(3, 1)$, by way of the isomorphism $\sigma$ that was described above. The matrix $\omega^\mu_\nu$ acts on the frame $\theta^\mu \wedge \theta^\nu$ by way of $\omega^\mu_\lambda \theta^\lambda \wedge \theta^\nu + \theta^\mu \wedge \omega^\nu_\lambda \theta^\lambda$, and on the components $F_{\mu\nu}$ by way of $-\omega^\lambda_\mu F_{\lambda\nu} - \omega^\lambda_\nu F_{\mu\lambda}$. .

Now that we have defined the action of the Lie algebra $\mathfrak{so}(3; \mathbf{C})$ on the frames of $SO(3; \mathbf{C})(\Lambda)$, the components of sections of $\Lambda^2(M)$ and the components of the metric tensor field on $\Lambda^2(M)$, we can introduce a connection on $SO(3; \mathbf{C})(\Lambda)$. We shall simply define such a connection by its 1-form. Hence, we now allow the infinitesimal complex rotation $\sigma^i_j$ to vary linearly with the tangent vectors to $SO(3; \mathbf{C})(\Lambda)$, or locally, with the tangent vectors to an open subset $U \subset M$ over which we have defined a local frame field $Z^i$. That is, a tangent vector $\mathbf{v} \in T_x(U)$ is linearly associated with an infinitesimal complex rotation $\sigma^i_j(\mathbf{v})$ of the frames in $SO(3; \mathbf{C})_x(\Lambda)$. One can then represent the 1-form $\sigma^i_j$ on $U$, which takes its values in $\mathfrak{so}(3; \mathbf{C})$, by a set of four functions $\Gamma^i_{j\mu}$, $\mu = 0, \ldots, 3$, on $U$ that also take their values in $\mathfrak{so}(3; \mathbf{C})$ by way of:

$$\sigma^i_j = \tilde{\Gamma}^i_{j\mu} \theta^\mu . \tag{6.1}$$

Note that in this form it is geometrically absurd to speak of permuting the lower two indices.

When $\mathbf{v}$ is the velocity vector field along some curve segment $C: [0, 1] \to U$ and $Y^i(\tau)$ is an oriented complex orthonormal frame field along $C$ one says that $Y^i$ is *parallel* along $C$ iff:

$$\mathbf{v} Y^i = i_\mathbf{v} D Y^i = \sigma^i_j(\mathbf{v}) Y^j . \tag{6.2}$$

If we define the *covariant differential* $\nabla Y^i$ of the frame field $Y^i$ by way of:

$$\nabla Y^i = DY^i - \sigma^i_j \otimes Y^j \tag{6.3}$$

then we can express the condition that $Y^i$ be parallel along the curve $C$ more concisely as:

$$\nabla_{\mathbf{v}} Y^i \equiv i_{\mathbf{v}} \nabla Y^i = 0 . \tag{6.4}$$

Hence, the covariant differential of a bivector field $\mathbf{F} = F^i Y_i$ is given by:

$$\nabla \mathbf{F} = dF^i \otimes Y_i + F^i \otimes \nabla Y_i = \nabla F^i \otimes Y_i , \tag{6.5}$$

with:

$$\nabla F^i = dF^i + \sigma^i_j F^j , \tag{6.6}$$

and the covariant differential of a 2-form $F = F_i Y^i$ is given by:

$$\nabla F = dF_i \otimes Y^i + F_i \otimes \nabla Y^i = \nabla F_i \otimes Y^i , \tag{6.7}$$

with:

$$\nabla F_i = dF_i - \sigma^i_j F^j . \tag{6.8}$$

If $V$ is a complex vector space on which $SO(3; \mathbf{C})$ acts linearly – i.e., a representation space for the group – then we can also define the *exterior covariant derivative* of a $V$-valued $k$-form $\alpha$ on $SO(3; \mathbf{C})(\Lambda)$ by the predictable rule:

$$\nabla \alpha = d\alpha + \sigma \wedge \alpha , \tag{6.9}$$

in which the meaning of the term $\sigma \wedge \alpha$ will again depend upon the way that $SO(3; \mathbf{C})(\Lambda)$ acts on $V$.

For instance, when one takes the exterior covariant derivatives of the canonical 2-form $Z^i$ and the connection 1-form $\sigma$ itself, one obtains the *Cartan structure equations*, which serve as the definitions of the torsion *3-form* $\Psi^i$ (N.B.) and the curvature 2-form $\Sigma^i_j$:

$$\Psi^i = \nabla Z^i = dZ^i + \sigma^i_j \wedge Z^i , \tag{6.10a}$$
$$\Sigma^i_j = \nabla \sigma^i_j = d\sigma^i_j + \sigma^i_k \wedge \sigma^k_j . \tag{6.10b}$$

This means that $\Psi^i$ takes its values in $\mathbf{C}^3$ and $\Sigma^i_j$ takes its values in $\mathfrak{so}(3; \mathbf{C})$.

A second application of the exterior covariant derivative to the structure equations gives the *Bianchi identities:*

$$\nabla \Psi^i = \nabla^2 Z^i = \Sigma^i_j \wedge Z^i , \tag{6.11a}$$
$$\nabla \Sigma^i_j = \nabla^2 \sigma^i_j = 0. \tag{6.11b}$$

On the bundle $SO(3; \mathbf{C})(\Lambda)$, we shall use the complex orthogonal metric $\gamma$ and the canonical 2-form $Z^i$ instead of the metric $g$ and the canonical 1-form $\theta^\mu$,

To say that the infinitesimal complex rotation that is described by the matrix $\sigma^i_k$ must preserve $\gamma$ is to say that:

$$\tilde{\nabla} \gamma_{ij} = d\gamma_{ij} - \sigma^i_k \gamma_{kj} - \sigma^i_k \gamma_{ik} = 0. \tag{6.12}$$

If the frame $Z^i$ is complex orthogonal then $\gamma_{ij} = \delta_{ij}$ and (6.12) takes the form:

$$\sigma_{ij} + \sigma_{ji} = 0 ; \tag{6.13}$$

i.e., the connection 1-form $\sigma^i_j$ must take its values in $\mathfrak{so}(3; \mathbf{C})$.

The condition that the connection $\sigma^i_j$ be torsionless then takes the form:

$$dZ^i = -\sigma^i_j \wedge Z^j. \tag{6.14}$$

The $\mathfrak{so}(3; \mathbf{C})$-connection on $SO(3; \mathbf{C})(\Lambda)$ that corresponds to the Levi-Civita connection on $SO_0(3, 1)(M)$ is also uniquely defined by the requirement that its torsion vanish:

$$\Psi^i = 0. \tag{6.15}$$

This makes the corresponding first Bianchi identity take the form:

$$0 = \Sigma^i_j \wedge Z^i. \tag{6.16}$$

We then define the Levi-Civita connection on $SO(3; \mathbf{C})(\Lambda)$ to be the unique $\mathfrak{so}(3; \mathbf{C})$ connection that has a vanishing torsion 3-form.

Now let us see how these conditions relate to the corresponding conditions on the Levi-Civita connection, as it is usually defined on $SO_0(3, 1)(M)$.

Since we have an isomorphism of the bundles $SO_0(3, 1)(M)$ and $SO(3; \mathbf{C})(\Lambda)$, as well as an isomorphism of the Lie algebra $\mathfrak{so}(3, 1)$ with $\mathfrak{so}(3; \mathbf{C})$, it stands to reason that we should also have an isomorphism of the affine space of $\mathfrak{so}(3, 1)$-connections on $SO_0(3, 1)(M)$ with the affine space of $\mathfrak{so}(3; \mathbf{C})$-connections on $SO(3, \mathbf{C})(\Lambda)$. To exhibit this one-to-one correspondence explicitly, we use the matrices $Z^i_{\mu\nu}$, $i = 1, 2, 3$, and their inverses $Z_i^{\mu\nu}$, $i = 1, 2, 3$, which relate a frame $Z_i$ on $\mathbf{C}^3$ to a bivector basis $\mathbf{e}_\mu \wedge \mathbf{e}_\nu$, $\mu < \nu$ by way of:

$$Z_i = \tfrac{1}{2} Z_i^{\mu\nu} \mathbf{e}_\mu \wedge \mathbf{e}_\nu. \tag{6.17}$$

We can then express the isomorphism of $\mathfrak{so}(3, 1)$ with $\mathfrak{so}(3; \mathbf{C})$ by way of:

$$\sigma_{jv}^{i\mu} = Z_{v\lambda}^i Z_j^{\lambda\mu}, \tag{6.18}$$

that is:

$$\sigma_j^i = Z_{\mu\lambda}^i Z_j^{\lambda v} \omega_v^\mu. \tag{6.19}$$

If $\omega_v^\mu$ is the 1-form for an $\mathfrak{so}(3, 1)$-connection on $SO_0(3, 1)(M)$ then one can associate it with a unique $\mathfrak{so}(3; \mathbf{C})$-connection $\sigma_j^i$ on $SO(3, \mathbf{C})(\Lambda)$ by pulling back $\omega_v^\mu$ to a 1-form on $SO(3, \mathbf{C})(\Lambda)$ by way of the bundle isomorphism and mapping the values of $\omega_v^\mu$ to $\mathfrak{so}(3; \mathbf{C})$ by way of the Lie algebra isomorphism. Briefly, in terms of local representatives for both 1-forms, this amounts to regarding equation (6.19) as involving 1-forms, instead of elements of the Lie algebras.

The first structure equation for $\omega_v^\mu$, namely:

$$\Theta^\mu = \nabla\theta^\mu = d\theta^\mu + \omega_v^\mu \wedge \theta^\mu, \tag{6.20}$$

can be related to the corresponding structure equation for $\sigma_j^i$ by exterior differentiating (4.2), keeping in mind that the $Z$-matrices are constant:

$$dZ^i = Z_{\mu v}^i d\theta^\mu \wedge \theta^v = Z_{\mu v}^i (\Theta^\mu - \omega_\lambda^\mu \theta^\lambda) \wedge \theta^v, \tag{6.21}$$

which gives:

$$Z_{\mu v}^i \Theta^\mu \wedge \theta^v = dZ^i + Z_{\mu v}^i \omega_\lambda^\mu \theta^\lambda \wedge \theta^v = \tilde{\nabla} Z^i \tag{6.22}$$

In order to make this consistent with (6.6a), we need only make the identification:

$$\Psi^i = Z_{\mu v}^i \Theta^\mu \wedge \theta^v. \tag{6.23}$$

This makes it clear that the representation of the torsion tensor for $\sigma_j^i$ is a set of 3-forms, not a set of 2-forms.

The second structure equation of $\omega$ maps to the second structure equation of $\sigma$ more directly, since $\sigma$ can be obtained by pulling back $\omega$ by means of the bundle isomorphism between $SO_0(3, 1)(M)$ and $SO(3; \mathbf{C})(\Lambda)$ and mapping the values of $\omega$ to $\mathfrak{so}(3; \mathbf{C})$ by means of the Lie algebra isomorphism between $\mathfrak{so}(3, 1)$ and $\mathfrak{so}(3; \mathbf{C})$. It helps that both pullbacks and the Lie algebra isomorphism commute with exterior derivatives.

We can expand the curvature 2-form $\Sigma_j^i$ in terms of the canonical 2-form $Z^i$:

$$\Sigma^i_j = \Sigma^i_{jk} Z^k .\tag{6.24}$$

If we want to relate this to the corresponding real form of the curvature then we can start with the locally equivalent expression:

$$\Sigma^i_j = \tfrac{1}{2} \Sigma^i_{jk} Z^k_{\mu\nu} \theta^\mu \wedge \theta^\mu .\tag{6.25}$$

in which we have expanded $Z^k$ in terms of real 2-forms, as in (3.51).

If we map $\mathfrak{so}(3; \mathbf{C})$ to $\mathfrak{so}(3, 1)$ by the isomorphism $\sigma$ then we can relate the components $R^\kappa{}_{\lambda\mu\nu}$ of the Riemannian curvature 2-form to those of its complex equivalent by way of:

$$R^\kappa{}_{\lambda\mu\nu} = \sigma^{j\kappa}_{i\lambda} \Sigma^i_{jk} Z^k_{\mu\nu} .\tag{6.26}$$

It is often more convenient to regard $\mathfrak{so}(3; \mathbf{C})$ as a Lie algebra over $\mathbf{C}^3$, instead of a matrix Lie algebra. As mentioned above in section 4, we can then use the same $Z^i$ matrices to map a real anti-symmetric 4×4 matrix $\omega_{\mu\nu}$ to a complex 3-vector in $\mathfrak{so}(3; \mathbf{C})$; in the case of the Riemannian curvature 2-form, one first lowers the upper index using $g$. As a result, one can also represent the 2-form $\Sigma^i_j$ with only one lower index; viz., $\Sigma_i$. At the risk of confusion, one can then expand the $\Sigma_i$ in terms of the $Z^i$:

$$\Sigma_i = \Sigma_{ij} Z^j = \tfrac{1}{2} \Sigma_{ij} Z^j_{\mu\nu} \theta^\mu \wedge \theta^\mu .\tag{6.27}$$

One can then relate the $\Sigma_{ij}$, which are now complex 3×3 matrices, to the $R_{\kappa\lambda\mu\nu}$ by way of:

$$\Sigma_{ij} = \tfrac{1}{4} Z^{\kappa\lambda}_i Z^{\mu\nu}_j R_{\kappa\lambda\mu\nu} .\tag{6.28}$$

We can expand this into:

$$\Sigma_{ij} = (R_{0i0j} - \frac{1}{4} \varepsilon_{ikl} \varepsilon_{jmn} R_{klmn}) + \frac{i}{2} \varepsilon_{jkl} R_{0ikl} .\tag{6.29}$$

in which all indices range from 1 to 3. This can be expressed in terms of 3×3 real matrices:

$$\Sigma_{ij} = \begin{bmatrix} R_{0101} & R_{0102} & R_{0103} \\ R_{0102} & R_{0202} & R_{0203} \\ R_{0103} & R_{0203} & R_{0303} \end{bmatrix} - \begin{bmatrix} R_{2323} & R_{2331} & R_{2312} \\ R_{2331} & R_{3131} & R_{3112} \\ R_{2312} & R_{3112} & R_{1212} \end{bmatrix} + i \begin{bmatrix} R_{0123} & R_{0131} & R_{0112} \\ R_{0131} & R_{0231} & R_{0212} \\ R_{0121} & R_{0212} & R_{0312} \end{bmatrix} .\tag{6.30}$$

The symmetry of these matrices follows from the basic symmetry of the Riemannian curvature tensor:

$$R_{\kappa\lambda\mu\nu} = R_{\mu\nu\kappa\lambda}. \tag{6.31}$$

Note that there is a conciseness in the way that the matrices in (6.30) represent the independent components of the Riemannian curvature tensor. The only redundancy is due to the symmetry of the matrices.

Since the matrix $\Sigma_{ij}$ is symmetric, one can either use it to define a second rank tensor on the bundle $SO(3; \mathbf{C})(\Lambda^2)$ via:

$$\Sigma = \Sigma_{ij} Z^i \otimes Z^j \tag{6.32}$$

or, similarly, another one on $\mathbf{C}^3$, if one regards the $Z^i$ as a complex 3-frame at a particular point of $M$. This second rank tensor is by no means non-degenerate in general, and, in fact, can very well be identically zero.

The group $SO(3; \mathbf{C})$ acts linearly on the components $\Sigma_{ij}$ since it also acts on the 3-frames $Z^i$. If $A^i_j \in SO(3; \mathbf{C})$ then the matrix $\Sigma_{ij}$ will go to $A^k_i A^l_j \Sigma_{kl}$, which can also be expressed as $A^T \Sigma A$. The action is reducible since, for one thing, $A^T = A^{-1}$ and the conjugation $A^{-1} \Sigma A$ will preserve the trace of $\Sigma$. Hence, as a first reduction, one can split $\Sigma$ into a traceless part and a trace:

$$\Sigma = \Sigma_0 + \tfrac{1}{3} \mathrm{Tr}(\Sigma) \gamma. \tag{6.33}$$

One finds, in fact, that:

$$\mathrm{Tr}(\Sigma) = \tfrac{1}{4} R. \tag{6.34}$$

This is related to the fact that:

$$R = [*(\theta^\mu \wedge \theta^\nu) \wedge \Omega_{\mu\nu}](\mathbf{V}) = (\theta^\mu \wedge \theta^\nu, \Omega_{\mu\nu}). \tag{6.35}$$

and the scalar product $(., .)$ is the real part of the complex scalar product $<.,.>_{\mathbf{C}}$.

There is a further reduction of the action of $SO(3; \mathbf{C})$ on the vector space of all $\Sigma$ that derives from the fact that $A$ is complex so it – or rather its representation as a 6×6 real matrix – must commute with *. Hence, $A$ will also take (anti-)self-dual 2-forms to other (anti-)self-dual 2-forms. This follows from the fact that if:

$$i*F = \pm F \tag{6.36}$$

then if $F' = A^T F A$ one must have that:

$$i*F' = i*A\mathrm{T}FA = A^T(i*F)A = \pm A\mathrm{T}FA = \pm F'. \tag{6.37}$$

Hence, we can decompose the traceless matrix $\Sigma_0$ into a self-dual part:

$$E = \tfrac{1}{2}(\Sigma_0 - i*\Sigma_0) \tag{6.38}$$

and an anti-self-dual part:

$$C = \tfrac{1}{2}(\Sigma_0 + i*\Sigma_0) \tag{6.39}$$

This is the standard Debever-Penrose decomposition of the Riemannian curvature tensor into the sum of the Weyl curvature $C$, the traceless part $E$ of the Ricci curvature tensor, and the scalar curvature $R$. The real form of $E$ has the components:

$$E_{\mu\nu} = R_{\mu\nu} - \tfrac{1}{4} R g_{\mu\nu}, \tag{6.40}$$

which differs from the Einstein tensor – viz., the divergenceless part of the Ricci tensor – by $1/4 R g_{\mu\nu}$. However, the vanishing of $E$ defines an Einstein space in the generalized sense of a Riemannian or pseudo-Riemannian manifold for which there exists a function $\alpha \in C(M)$ such that:

$$R_{\mu\nu} = \alpha g_{\mu\nu}, \tag{6.41}$$

which includes the cases $\alpha = 0$, $1/4R$, and $1/2R$.

Hence, one arrives at the traditional way of expressing the vacuum Einstein field equations in complex form: the vanishing of the self-dual part of the curvature 2-form $\Sigma$.

As for the anti-self-dual tensor $C$, i.e., the Weyl tensor, it defines a quadratic form on $\mathbf{C}^3$:

$$C(\mathbf{X}, \mathbf{X}) = C_{ij} X^i X^j. \tag{6.42}$$

The zero hypersurface of this quadratic form defines a quadric surface in $\mathbf{C}^3$, and, since its equation is homogeneous of degree two, a quadric surface in $\mathbf{CP}^2$.

We already have another quadric defined in $\mathbf{CP}^2$ by way of the complex orthogonal structure:

$$0 = \gamma(\mathbf{X}, \mathbf{X}) = \gamma_{ij} X^i X^j. \tag{6.43}$$

The four intersection points of these two quadrics define the basis for the Petrov classification of Weyl tensors, as described in Debever [**2, 3**]. However, since we have not fundamentally altered anything in that analysis except to represent $\mathbf{CP}^2$ directly in terms of the duality planes in $A^2$ instead of something derived from projecting the complex vector space of self-dual 2-forms – a space that is also isometric to $\mathbf{C}^3$ with the standard Euclidian structure – we shall not elaborate on the details.

## 7 Discussion

Although it may seem that the representation of general relativity in terms of the bundle of 2-forms on spacetime instead of the tangent bundle is reasonably complete, concise, and self-contained, nevertheless, there are a few topics of a physical nature that also suggest directions of further research if one is to make the representation truly complete.

For one thing, it is usually only the vacuum Einstein equations that are discussed in the context of complex relativity. In order to discuss the field equations for gravitation with sources, one must also represent the energy-momentum-stress tensor in terms of things that are germane to the bundle of 2-forms. Since the physical concept of energy-momentum is more intrinsically related to tangent vectors and covectors associated with motion, this would involve a non-trivial resetting of mechanics in general.

This brings us to the second issue: Although we have discussed the complex geometric equivalent of the Einstein field equations, we have said nothing about what happens to the geodesic equations of motion. Indeed, one sees that since geodesics relate to the parallel translation of tangent vectors along curves, there is a fundamental issue associated with how one does something similar with the geometric objects that relate to the bundle of bivectors and 2-forms in place of the tangent and cotangent bundles. However, it seems plausible to conjecture that perhaps the type of motion that is intrinsically related to fields of 2-planes is wave motion, if one assumes that the 2-planes are tangent to the momentary wavefronts, such as the polarization planes of electromagnetic waves. Conceivably, one might obtain these surfaces as autoparallel surfaces for a connection that defines the parallel translation of 2-planes, or, equivalently, 2-forms. The motion of the momentary wave front could then come about from the parallel translation of the 2-planes along the dual surfaces that have the dual 2-planes for their tangent planes. This also suggests an immediate dichotomy between the decomposable 2-forms $F$ such that $F \wedge {}^*F$ is non-zero and the ones for which it vanishes. Since it is the latter class that defines lightlike lines in the tangent spaces, one speculates that the difference between the two types of 2-forms is one of characteristic versus non-characteristic wave motion. Perhaps the characteristic wave motion, which is more closely related to the complex structure on $\Lambda^2(M)$ defined by $*$, represents, in some sense, a "complexification" of geodesic motion from the parallel translation of real tangent lines along real curves to the parallel translation of complex lines along complex curves.

One of the predictable directions of inquiry in complex relativity that followed the study of self-duality in Euclidian gauge field theories was the definition and study of gravitational instantons, which would presumably be analogous, in some sense, to the instanton solutions of the gauge field theories. However, since the application of self-duality to gauge field theories was, as was pointed out, mostly carried out in the context of Riemannian manifolds, not surprisingly, the analogous application to gravitation was in the context of Euclidian quantum gravity. Since Euclidian **R**$^4$ and Minkowski space both live in the same complex orthogonal space, presumably what happens in one of them can be "Wick rotated" into the other. However, there are limits to this logic, since the theory of the Laplace equation is not completely isomorphic to the theory of the wave equation in all of its aspects, even though one can generalize the Laplace operator to pseudo-Riemannian manifolds of unspecified signature type. This has, as a consequence, that Hodge theory is not applicable to Lorentzian manifolds. Nevertheless, since gravitational instantons represent an important class of solutions to the Einstein equations, the details of how one represents them in the presently discussed formalism is worth investigating.

Another topic of general relativity that is closely related to the methods of complex relativity is the method of the Ashtekar variables, which are complex variables that define a phase space for the Hamiltonian formulation of gravitation as a problem in the time evolution of spatial geometry, or "geometrodynamics." Although the definitions of the variables start in the complexification of Minkowski space, it is possible that they too can be represented in the present framework.